\documentclass[aps,pra,showpacs,amsmath,amssymb]{revtex4-1}
\usepackage{graphicx} 

\begin{document}

\title{Slow polaritons with orbital angular momentum in atomic gases}

\author{J.~Ruseckas, A.~Mekys and G.~Juzeli\=unas}

\affiliation{Institute of Theoretical Physics and Astronomy,
Vilnius University,\\ A. Go\v{s}tauto 12, LT-01108 Vilnius, Lithuania}

\date{\today{}}

\begin{abstract}
Polariton formalism is applied for studying the propagation of a probe field of
light in a cloud of cold atoms influenced by two control laser beams of larger
intensity. The laser beams couple resonantly three hyperfine atomic ground
states to a common excited state thus forming a tripod configuration of the
atomic energy levels involved. The first control beam can have an optical
vortex with the intensity of the beam going to zero at the vortex core. The second 
control beam without a vortex ensures the loseless (adiabatic) propagation of the 
probe beam at a vortex core of the first control laser. We investigate the storage of
the probe pulse into atomic coherences by switching off the control beams, as
well as its subsequent retrieval by switching the control beams on. The optical vortex is
transferred from the control to the probe fields during the storage or
retrieval of the probe field. We analyze conditions for the vortex to be transferred efficiently
to the regenerated probe beam and discuss possibilities of experimental implementation
of the proposed scheme using atoms like rubidium or sodium. 
\end{abstract}

\pacs{42.50.Gy,03.67.-a,42.50.Tx}

\maketitle

\section{Introduction}

\begin{figure}
\includegraphics[width=0.6\textwidth]{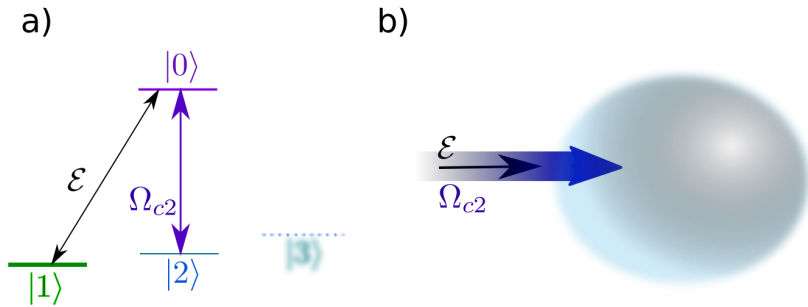}
\caption{Probe and control laser beams acting on atoms characterized by two hyperfine 
ground states $|1\rangle$ and $|2\rangle$, as well as an 
excited state $|0\rangle$ to form a three-level scheme of the $\Lambda$ type. 
Atoms are initially in the ground state $|1\rangle$  . Stimulated exchange of photons between 
the probe ($\mathcal{E}$) and contol ($\Omega_{c2}$) laser fields 
creates a superposition of the hyperfine atomic ground states $|1\rangle$ and $|2\rangle$ 
making the medium transparent for the resonant probe pulse. }
\label{fig:fig1-Lambda}
\end{figure}

During the last several years there has been a great deal of interest in slow
\cite{Hau-99,Kash99,Budker99,Novikova07PRL,Davidson09NP}, stored
\cite{Fleischhauer-00,Liu-01,Phill2001,Juzeliunas-02,Scully02PRL,Yanik05PRA,Gor05EPJD,Lukin05Nature,Ite07PRA,Hau07Nature,Bloch09PRL,Hau09PRL,Beil10PRA} and
stationary
\cite{Stationary-light-03--09,Stationary-light2,Stationary-light4,Stationary-ligth3,Stationary-ligth5,Otterbach10PRL,Unanyan10PRL} light. Light can be slowed down by seven
orders of magnitude to velocities of several of tens of meters per second
\cite{Hau-99} due to the electromagnetically induced transparency (EIT)
\cite{Arimondo-96,Harris-97,Scully-97,Lukin-03,Fleischhauer-05}. The EIT makes
a resonant and opaque medium transparent for a probe beam by applying a control
laser beam of larger intensity. The probe beam couples resonantly the ground
and excited atomic states, whereas the control beam couples the same excited
state to another unpopulated atomic ground state. This makes a $\Lambda$
configuration of the atomic energy levels involved, as depicted in
Fig.~\ref{fig:fig1-Lambda}.
The optical transitions induced by both laser beams interfere destructively
preventing population of the excited atomic state. As a result, a weak pulse of
probe light travels slowly and with little losses in a resonant medium due to
the application of the control laser beam.

The EIT was shown not only to slow down dramatically light pulses
\cite{Hau-99,Kash99,Budker99,Novikova07PRL,Davidson09NP}, but also to store
them
\cite{Liu-01,Phill2001,Lukin05Nature,Hau07Nature,Bloch09PRL,Hau09PRL,Beil10PRA}
in atomic gases. The storage and release of a probe pulse has been accomplished
\cite{Liu-01,Phill2001,Lukin05Nature,Hau07Nature,Bloch09PRL,Hau09PRL,Beil10PRA}
by switching off and on the control laser \cite{Fleischhauer-00}. The coherent
control of the propagation of quantum light pulses can lead to a number of
applications, such as generation of non-classical states in atomic ensembles
and reversible quantum memories for slow light
\cite{Fleischhauer-00,Juzeliunas-02,Scully02PRL,Lukin05Nature,Lukin-03,Fleischhauer-05,Appel08PRL,Honda08PRL,Akiba09NJP}. On the other hand, propagation of
slow light through moving media
\cite{Leonhardt00PRL,Ohberg02PRAR,Fleischhauer-Gong-PRL02,Juz-moving,Artoni03PRA,Zimmer-Fleischhauer-PRL04,Padgett-06,Ruseckas-07} may be used for the light
memories and rotational sensing devices.

\begin{figure}
\includegraphics[width=0.9\textwidth]{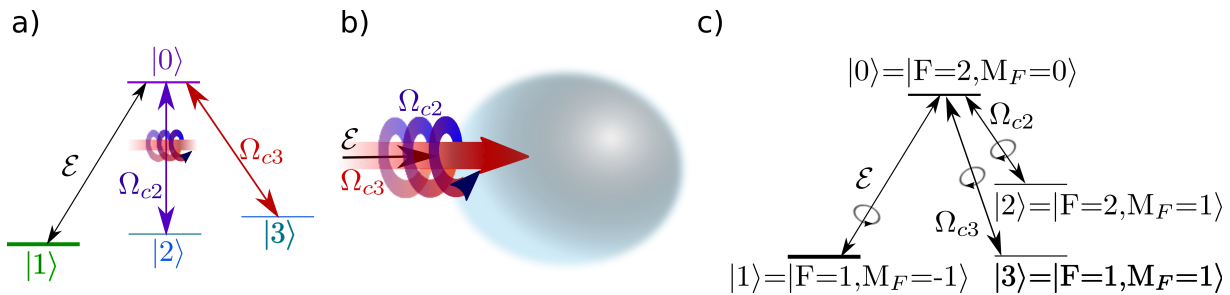}
\caption{(a) Tripod scheme of the atom-light coupling involving a probe beams ($\mathcal{E}$)  and
two control beams ($\Omega_{c2}$ and $\Omega_{c3}$). The three beams induce transitions between 
the atomic excited state $|0\rangle$ and three ground states
 $|1\rangle$, $|2\rangle$ and  $|3\rangle$. (a), (b) A control beam with the Rabi frequency 
 $\Omega_{c2}$ can have an optical vortex.  Application of an additional control laser beam
without the OAM ($\Omega_{c3}\ne0$) makes it possible to avoid losses in the
propagation of the probe beam at the vortex core where the amplitude
$\Omega_{c2}$ goes to zero. (c) A possible experimental realization 
of the tripod setup for atoms like Sodium \cite{Liu-01} or Rubidium  \cite{Phill2001}
containing the hyperfine ground states with $F=1$ and  $F=2$.}
\label{fig:fig1}
\end{figure}

The orbital angular momentum (OAM) \cite{Allen-99,Allen-03} provides a new
element to the slow light giving additional possibilities in manipulation of the optical information during the storage and retrieval of the slow light. The previous studies have concentrated on situations where the probe beam contains an OAM
\cite{Ruost04PRL,Davidson07PRL,Ruseckas-07,Yelin08PRA,Moretti09PRA}. In the
present paper we consider another scenario in which it is a control laser beam
which can carry an optical vortex. The intensity of such a control beam goes to
zero at the vortex core leading to the absorption losses of the probe beam in
this area. To avoid the losses we suggest to use an additional control laser
without an optical vortex, so that the total intensity of the control lasers is
non-zero at the vortex core of the first control laser. The probe and both
control laser fields induce transitions between the atomic energy levels in a
tripod configuration of the light-atom coupling
\cite{Unanian,Knight-02,Rebic04PRA,Petrosyan04PRA,Yelin04PRA,Rus05,Mazets05PRA,Zaremba06OC,Zaremba07PRA,Zaremba07OC,Garva07OC,RusecMekJuz08} as depicted in
Fig.~\ref{fig:fig1}a. We show that the regenerated slow light can acquire the
OAM if one of the control beams contains it. We explore conditions for the
optical vortex of the control beam to be transferred efficiently to the
regenerated probe beam. 

The tripod scheme can be realised for atoms like Sodium \cite{Liu-01} 
or Rubidium  \cite{Phill2001} containing two hyperfine ground levels with $F=1$ and $F=2$,
as depicted in Fig.~\ref{fig:fig1}c. These atoms have been employed in the original experiments on the
storage of slow light based on a simpler $\Lambda$ setup \cite{Liu-01,Phill2001}. 
In the present situation  $|1\rangle$ and $|3\rangle$ 
correspond to the magnetic sublevels (with $m_F =1$ and 
$m_F =-1$) of the $F=1$ hyperfine ground level, whereas the state  $|3\rangle$ represents the 
hyperfine ground state with $F=2$ and  $m_F=1$. The probe beam 
is to be $\sigma^{+}$ polarized, whereas both control beams 
are to be $\sigma^{-}$ polarized to make a tripod setup. 
Such a scheme can be produced by adding an extra 
circularly polarized laser beam $\Omega_{c3}$ as compared to the experiment 
by Liu et al \cite{Liu-01} on the light storage in the sodium gases using the $\Lambda$ scheme. 
Thus it is feasible  to implement the suggested experiment on the transfer 
of optical vortex from the control to the probe fields using the tripod setup.

The storage and retrieval of slow light is analyzed using the polariton
formalism. The starting point is a set of the atomic equations together with
the equation for the probe field. Subsequently we obtain two coupled equations
for dark-state polaritons representing the slow light in the atomic medium. We
provide conditions when the polaritons are decoupled. An advantage of polariton
formalism is a simplicity of the relationship between the polariton field and
the regenerated electric field, a feature which is missing in the direct
analysis of the probe beam propagation \cite{RusecMekJuz08}. Furthermore the
equation for the polariton has a usual form of matter wave equation which
describes the atomic evolution when the control fields are off.

\section{Initial equations}

We will deal with an ensemble of atoms characterized by three hyperfine ground
states $|1\rangle$, $|2\rangle$, and $|3\rangle$, as well as an electronic
excited state $|0\rangle$. The atomic internal and center of mass dynamics is
represented by a four component field $\Psi(\mathbf{r})$. Its components
$\Psi_1(\mathbf{r},t)$, $\Psi_2(\mathbf{r},t)$, $\Psi_3(\mathbf{r},t)$, and
$\Psi_0(\mathbf{r},t)$ describe the atomic center of mass motion in the
corresponding internal states $|1\rangle$, $|2\rangle$, $|3\rangle$ and
$|0\rangle$. In the semiclassical (mean field) approach, $\Psi_j(\mathbf{r},t)$
defines the probability amplitude to find an atom positioned at $\mathbf{r}$ in
the $j$-th internal state, with $j=0,1,2,3$. In the fully quantum approach
$\Psi_j(\mathbf{r},t)$ is the corresponding field operator.

Three beams of light act on the atoms in a tripod configuration of the
atom-light coupling
\cite{Unanian,Knight-02,Rebic04PRA,Petrosyan04PRA,Yelin04PRA,Rus05,Mazets05PRA,Zaremba06OC,Zaremba07PRA,Zaremba07OC,Garva07OC,RusecMekJuz08}. Two strong classical control
lasers induce transitions $|2\rangle\rightarrow|0\rangle$ and
$|3\rangle\rightarrow|0\rangle$, whereas a weaker probe field drives a
transition $|1\rangle\rightarrow|0\rangle$, as shown in Fig.~\ref{fig:fig1}.
The former control lasers are characterised by the Rabi frequencies
$\Omega_{c2}$ and $\Omega_{c3}$ to be treated as incident variables. The latter
probe beam is a dynamical quantity described by the electric field strength 
\begin{equation}
\mathbf{E}(\mathbf{r},t)=\hat{\mathbf{e}}\sqrt{\frac{\hbar\omega}{
2\varepsilon_0}}\mathcal{E}(\mathbf{r},t)e^{-i\omega t}+\mathrm{H.c.}\,,
\end{equation}
where $\omega=ck$ is the central frequency of the probe photons,
$\mathbf{k}=\hat{\mathbf{z}}k$ is the wave vector, and
$\hat{\mathbf{e}}\bot\hat{\mathbf{z}}$ is the unit polarization vector. The
probe field can be treated either as a classical variable or as a quantum
operator. The dimensions of the electric field amplitude $\mathcal{E}$ are
chosen such that its squared modulus represents a number density of probe
photons.

The probe field is considered to be quasi-monochromatic, and its amplitude
$\mathcal{E}\equiv\mathcal{E}(\mathbf{r},t)$ changes little over the optical
cycle. The slowly (in time) varying amplitude of the probe field obeys the
following equation: 
\begin{equation}
\left(\frac{\partial}{\partial
t}-i\frac{c^2}{2\omega}\nabla^2-i\frac{\omega}{2}\right)\mathcal{
E}=ig\Phi_1^*\Phi_0\,,
\label{eq:electric}
\end{equation}
where the parameter $g=\mu\sqrt{\omega/2\varepsilon_0\hbar}$ characterizes the
strength of coupling of the probe field with the atoms, $\mu$ being the dipole
moment of the atomic transition $|1\rangle\rightarrow|0\rangle$. The quantities
on the r.h.s.\ of Eq.~(\ref{eq:electric}) $\Phi_0$ and $\Phi_1^*$ represent
slowly (in time) varying atomic fields. The asterisk in $\Phi_1^*$ refers
either to the complex conjugation of a classical field or the Hermitian
conjugation of a quantum field. The slowly varying atomic fields $\Phi_j$
($j=1,2,3,4$) are related to the original ones as:
$\Phi_1=\Psi_1e^{i\omega_1t}$,
$\Phi_2=\Psi_2e^{i(\omega_1+\omega-\omega_{c2})t}$,
$\Phi_3=\Psi_3e^{i(\omega_1+\omega-\omega_{c3})t},$
$\Phi_0=\Psi_0e^{i(\omega_1+\omega)t}$, where $\hbar\omega_1$ is the energy of
the atomic ground state $1$, whereas $\omega_{c2}$ and $\omega_{c3}$ are the
frequencies of the control fields.

The atomic equations of motion read
\begin{eqnarray}
\hat{K}\Phi_1 & = & V_1(\mathbf{r})\Phi_1-\hbar
g\mathcal{E}^*\Phi_0\,,
\label{eq:phi1}
\\\hat{K}\Phi_0 & = &
\hbar(\omega_{01}-i\gamma)\Phi_0+V_0(\mathbf{r})\Phi_0-\hbar\Omega_{c2}\Phi_2
-\hbar\Omega_{c3}\Phi_3-\hbar g\mathcal{E}\Phi_1\,,
\label{eq:phi0}
\\
\hat{K}\Phi_2 & = &
\hbar\omega_{21}\Phi_2+V_2(\mathbf{r})\Phi_2-\hbar\Omega_{c2}^*\Phi_0\,,
\label{eq:phi2}
\\\hat{K}\Phi_3 & = &
\hbar\omega_{31}\Phi_3+V_3(\mathbf{r})\Phi_3-\hbar\Omega_{c3}^*\Phi_0\,,
\label{eq:phi3}
\end{eqnarray}
with
\begin{equation}
\hat{K}=i\hbar\frac{\partial}{\partial t}+\frac{\hbar^2}{2m}\nabla^2\,,
\end{equation}
where $\omega_{21}=\omega_2-\omega_1+\omega_{c2}-\omega$ and
$\omega_{31}=\omega_3-\omega_1+\omega_{c3}-\omega$ are the frequencies of the
electronic detuning from the two-photon resonances,
$\omega_{01}=\omega_0-\omega_1-\omega$ is the frequency of the electronic
detuning from the one-photon resonance, and $\gamma$ is the decay rate of the
excited electronic level. Note that the inclusion of the non-zero decay rates
should be generally accompanied by introduction of the noise operator in the
equations of motion \cite{Scully-97}. Yet in the present situation one can
disregards the latter noise: we are working in the linear regime with respect
to the probe field, so the population of the excited state is small enough.
Here also $m$ is the atomic mass and $V_j(\mathbf{r})$ is the trapping
potential for an atom in the internal state $j$ ($j=1,2,3,0$). The terms
containing atomic mass $m$ are important for the description of the
light-dragging effects
\cite{Leonhardt00PRL,Ohberg02PRAR,Fleischhauer-Gong-PRL02,Juz-moving,Artoni03PRA,Zimmer-Fleischhauer-PRL04,Padgett-06,Ruseckas-07}.

In Eqs.~(\ref{eq:phi1})--(\ref{eq:phi3}) the coupling of atoms with the probe
and control fields has been written using the rotating wave approximation.
Therefore, the last term in Eq.~(\ref{eq:phi1}) has a negative frequency part
of the probe field ($\mathcal{E}^*$), whereas the last term in
Eq.~(\ref{eq:phi0}) has a positive frequency part ($\mathcal{E}$). Similarly
Eq.~(\ref{eq:phi0}) contains Rabi frequencies $\Omega_{c2}$ and $\Omega_{c3}$,
whereas Eqs.~(\ref{eq:phi2}) and (\ref{eq:phi3}) contain their complex
conjugated counterparts $\Omega_{c2}^*$ and $\Omega_{c3}^*$.

The equation of motion (\ref{eq:phi1}) for $\Phi_1$ does not explicitly
accommodate collisions between the ground-state atoms. If the atoms in the
internal ground-state $1$ form a Bose-Einstein condensate (BEC), the
collisional effects can be included replacing $V_1(\mathbf{r})$ by
$V_1(\mathbf{r})+g_{11}|\Phi_1|^2$ in Eq.~(\ref{eq:phi1}), where
$g_{11}=4\pi\hbar^2a_{11}/m$ and $a_{11}$ is the scattering length between the
condensate atoms in the internal state $1$. This yields a mean-field equation
for the condensate wave function $\Phi_1$.

Initially the atoms populate the ground level $1$. We are interested in the
linear regime where the modulus of Rabi frequency of the probe field
$\Omega_p=g\mathcal{E}$ is much smaller than the total Rabi frequency of the
control beams $\Omega_c$ given by Eq.~(\ref{eq:Omega-c}) below. Consequently
one can neglect the last term in Eq.~(\ref{eq:phi1}) that causes depletion of
the ground level $1$. This provides a closed equation for the ground state
dynamics: $\hat{K}\Phi_1=V_1(\mathbf{r})\Phi_1.$ If the atoms in the internal
ground-state $1$ form a BEC, its wave function $\Phi_1=\sqrt{n}\exp(iS_1)$
represents an incident variable determining the atomic density $n$ and the
condensate phase $S_1$. The latter phase will not play an important role in our
subsequent analysis, since we are not interested in the influence of the
condensate dynamics on the propagation of slow light. The phase will be taken
to be zero ($S_1=0$) when dealing with the storage and retrieval of slow light
in the Section \ref{sec:Storage-and-release}.

\section{Dark- and bright-state polaritons}

When the probe photons enter the atomic media, they are converted into
composite quasiparticles of the radiation and atomic excitations known as
polaritons. Let us first introduce the bright-state polariton $\Phi_B$: 
\begin{equation}
\Phi_B=\zeta_c(\xi_{c2}\Phi_2+\xi_{c3}\Phi_3)+\zeta_1\mathcal{E}\,,
\label{eq:brightP}
\end{equation}
where
\begin{equation}
\Omega_c=\sqrt{|\Omega_{c2}|^2+|\Omega_{c3}|^2}
\label{eq:Omega-c}
\end{equation}
is the total Rabi frequency, 
\begin{eqnarray}
\xi_{c2} & = &
\Omega_{c2}/\Omega_c\,,\qquad\xi_{c3}=\Omega_{c3}/\Omega_c\,,
\label{eq:zeta-2--3}
\\\zeta_1 & = &
g\Phi_1/\Xi\,,\qquad\zeta_c=\Omega_c/\Xi\,
\label{eq:zeta-1}
\end{eqnarray}
are dimensionless parameters, and
\begin{equation}
\Xi=\sqrt{\Omega_c^2+g^2n}\,.
\end{equation}
The polariton $\Phi_B$ represents a specific superposition of the atomic and
the probe fields featured in the equation of motion (\ref{eq:phi0}) for atoms
in the excited electronic state. The latter equation (\ref{eq:phi0}) can be
rewritten in terms of the bright-state polariton 
\begin{equation}
\hat{K}\Phi_0=\hbar(\omega_{01}-i\gamma)\Phi_0+V_0(\mathbf{r})\Phi_0
-\hbar\Xi\Phi_B\,.
\label{eq:phi-0-altern}
\end{equation}
In this way the bright-state polariton is responsible for the light-induced
atomic transitions to the excited state.

The two dark-state polaritons are defined as superpositions of the atomic
coherences and the probe photons orthogonal to the bright-state polariton
$\Phi_{B1}$:
\begin{eqnarray}
\Phi_{D1} & = &
\zeta_c\mathcal{E}-\zeta_1^*(\xi_{c2}\Phi_2+\xi_{c3}\Phi_3)\,,
\label{eq:darkP1}
\\\Phi_{D2} & = &
\xi_{c3}^*\Phi_2-\xi_{c2}^*\Phi_3\,.
\label{eq:darkP2}
\end{eqnarray}
It is to be noted that only the first dark-state polariton $\Phi_{D1}$ of the
tripod scheme contains the electric probe field component and thus has a
non-zero radiative group velocity. The incoming light is converted exclusively
into this polariton when it enters the medium. The second dark-state polariton
$\Phi_{D2}$ does not have any contribution by the probe photons and is thus
characterised by a zero radiative group velocity. It corresponds to the dark
state of the $\Lambda$ system consisting of the levels $2$, $3$, and $0$. The
combination $\xi_{c2}\Phi_2+\xi_{c3}\Phi_3$ featured in Eqs.~(\ref{eq:brightP})
and (\ref{eq:darkP1}), represents the bright state of such a $\Lambda$ system.
In this way, only the first polariton experiences the radiative motion, the
second one being trapped in the atomic medium.

The ``bare'' atomic and probe fields can be cast in terms of the dark and
bright polaritons of the tripod system as:
\begin{eqnarray}
\Phi_2 & = &
\xi_{c2}^*(\zeta_c\Phi_B-\zeta_1\Phi_{D1})+\xi_{c3}\Phi_{D2}
\label{eq:phi2-DB}
\\\Phi_3 & = &
\xi_{c3}^*(\zeta_c\Phi_B-\zeta_1\Phi_{D1})-\xi_{c2}\Phi_{D2}
\label{eq:phi3-DB}
\\\mathcal{E} & = &
\zeta_1^*\Phi_B+\zeta_c\Phi_{D1}
\label{eq:electric-DB}
\end{eqnarray}
To obtain the equation for the dark-state polaritons one needs to take the time
derivative of Eqs.~(\ref{eq:darkP1})--(\ref{eq:darkP2}) and make use of the
equations of motion (\ref{eq:electric}), (\ref{eq:phi2})--(\ref{eq:phi3}) and
(\ref{eq:phi-0-altern}).

Suppose the control and probe beams are tuned close to the two-photon
resonance. Application of such beams cause EIT in which the transitions
$|1\rangle\rightarrow|0\rangle$, $|2\rangle\rightarrow|0\rangle$, and
$|3\rangle\rightarrow|0\rangle$ interfere destructively preventing population
of the excited state $0$. As a result, the atom-light system is driven to the
dark states, and the bright state polariton $\Phi_B$ (featured in the equation
of motion (\ref{eq:phi-0-altern}) for the excited state atoms) is weakly
populated: $\Phi_B\approx0$. Neglecting the contribution due to the
bright-state polariton $\Phi_B$ (adiabatic approximation), one obtains the
equations for the dark-state polaritons $\Phi_{D1}$ and $\Phi_{D2}$.
Introducing a column $\Phi=(\Phi_{D1},\Phi_{D2})^T$, it is convenient to
represent these equations in a matrix form:
\begin{equation}
i\hbar\frac{\partial}{\partial t}\Phi=\left[-\frac{\hbar^2}{2}\left(
\begin{array}{cc}
1/m_{D1} & 0\\ 0 & 1/m\end{array}\right)\nabla^2+i\hbar\mathbf{J}\cdot\nabla
+U\right]\Phi\,,
\label{eq:polar1}
\end{equation}
where the $2\times2$ matrices $\mathbf{J}$ and $U$ are defined in the
Appendix~\ref{sec:appA}. The former $\mathbf{J}$ represents a complex vector
potential, $U$ being a complex scalar potential. Even though the potentials are
complex, the equation of motion (\ref{eq:polar1}) is Hermitian and thus it
preserves the norm of the column $\Phi$. Here also
\begin{equation}
m_{D1}=\left(\frac{c^2}{\hbar\omega}|\zeta_c|^2+\frac{1}{m}|\zeta_1|^2\right)^{
-1}
\label{eq:mD1}
\end{equation}
is the effective mass of the first dark-state polariton. The mass $m_{D1}$
exhibits position- and time-dependence through its dependence on the Rabi
frequencies of the control fields and also on the atomic density. The second
polaritons does not have a radiative component, so its effective mass coincides
with the atomic mass $m$ in Eq.~(\ref{eq:polar1}).

The effective mass of the first polariton can be represented as: 
\begin{equation}
m_{D1}=\left(\frac{1}{m_{\mathrm{rad}}}+\frac{1}{m}\frac{g^2n}{\Omega_c^2
+g^2n}\right)^{-1}\,,
\label{eq:mD1-a}
\end{equation}
where

\begin{equation}
m_{\mathrm{rad}}=\frac{\hbar\omega}{cv_{\mathrm{rad}}}=m\frac{v_{\mathrm{rec}}}{
v_{\mathrm{rad}}}
\label{eq:m-rad}
\end{equation}
and
\begin{equation}
v_{\mathrm{rad}}=\frac{c^2\Omega_c^2}{\Omega_c^2+g^2n}
\label{eq:v-rad}
\end{equation}
are, respectively, the radiative {}``mass{}`` and the radiative group velocity
of the first polariton, $v_{\mathrm{rec}}=\hbar\omega/mc$ being the atomic
recoil velocity. In the slow light regime where $\Omega_c^2\ll g^2n$, the
latter $v_{\mathrm{rad}}\approx c^2\Omega_c^2/g^2n$ is much smaller than the
vacuum speed of light: $v_{\mathrm{rad}}\ll c$. The radiative velocity
$v_{\mathrm{rad}}$ can be of the order of $10\,\mathrm{m/s}$ for the slow light
in atomic gases \cite{Hau-99}. This greatly exceeds the typical velocities
associated with the centre of mass motion of cold atoms. For instance, the
atomic recoil velocity is typically of the order of $1\,\mathrm{cm/s}$. Thus
the second term can be neglected in Eq.~(\ref{eq:mD1-a}), giving $m_{D1}\approx
m_{\mathrm{rad}}$.

\subsection{Co-propagating probe and control beams}

Suppose that the control beams propagate along $z$ axis with $k_{c2}\approx
k_{c3}=k_c$:
\begin{equation}
\Omega_{c2}=\Omega_{c2}^{\prime}e^{ik_cz}\,,\qquad\Omega_{c3}=\Omega_{c3}^{
\prime}e^{ik_cz}\,.
\label{eq:Omega-c2--c3}
\end{equation}
For paraxial control beams the amplitudes $\Omega_{c2}^{\prime}$ and
$\Omega_{c3}^{\prime}$ depend weakly on the propagation direction $z$. It is
convenient to represent the dark-state polaritons as:
\begin{eqnarray}
\Phi_{D1}(\mathbf{r},t) & = &\Phi_{D1}^{\prime}(\mathbf{r},t)e^{ikz}\,,\\
\Phi_{D2}(\mathbf{r},t) & = &\Phi_{D2}^{\prime}(\mathbf{r},t)e^{-ikz}\,,
\end{eqnarray}
where the amplitudes $\Phi_{D1}^{\prime}(\mathbf{r},t)$ and
$\Phi_{D2}^{\prime}(\mathbf{r},t)$ depend slowly on the propagation direction
$z$ in the paraxial case. Introducing a column
$\Phi^{\prime}=(\Phi_{D1}^{\prime},\Phi_{D2}^{\prime})^T$,
Eq.~(\ref{eq:polar1}), provides the following equation for the slowly varying
amplitudes:
\begin{equation}
i\hbar\left[\frac{\partial}{\partial t}+\left(
\begin{array}{cc}
v_{g1} & 0\\ 0 & 0
\end{array}\right)\frac{\partial}{\partial z}\right]\Phi^{\prime}=\left[-\frac{
\hbar^2}{2}\left(
\begin{array}{cc}
\frac{1}{m_{D1}} & 0\\ 0 &\frac{1}{m}\end{array}\right)\nabla^2+i\hbar\mathbf{
J}^{\prime}\cdot\nabla+U^{\prime}\right]\Phi^{\prime}\,,
\label{eq:polar2}
\end{equation}
where the $2\times2$ matrices $\mathbf{J}^{\prime}$ and $U^{\prime}$ are
presented in the Appendix~\ref{sec:appB}. Here
\begin{equation}
v_{g1}=v_{\mathrm{rad}}+\frac{\hbar}{m}(k-k_c)|\zeta_1|^2
\label{eq:v-g1}
\end{equation}
is the group velocity of the first dark-state polariton. It comprises the
radiative group velocity and the velocity of the two photon recoil. The latter
term can be neglected giving $v_{g1}\approx v_{\mathrm{rad}}$.

\subsection{Decoupled dark-state polaritons}

Let us analyse the terms which couple both dark polaritons in the equation of
motion (\ref{eq:polar2}). The term with time derivatives in the non-diagonal
elements of the matrix $U^{\prime}$ is proportional to
\begin{equation}
\xi_{c2}\frac{\partial}{\partial t}\xi_{c3}-\xi_{c3}\frac{\partial}{\partial
t}\xi_{c2}=\frac{\Omega_{c2}}{\Omega_c^2}\frac{\partial}{\partial
t}\Omega_{c3}-\frac{\Omega_{c3}}{\Omega_c^2}\frac{\partial}{\partial
t}\Omega_{c2}\,.
\end{equation}
If both control pulses depend on time in the same manner, i.e.\
$\Omega_{c2}=\Omega_{c2}^{(0)}f(t)$ and $\Omega_{c3}=\Omega_{c3}^{(0)}f(t)$,
the above term is zero. Thus the coupling between the two dark-state polaritons
can be avoided by switching both control pulses off and on in the same way, so
that both of them exhibit the same temporal behaviour.

Let us next estimate non-diagonal terms which contain the spatial derivatives
of the control pulses in the equation of motion (\ref{eq:polar2}) and hence
couple both dark polaritons. Such non-diagonal matrix elements are of the order
of the atomic recoil energy $\hbar\omega_{\mathrm{rec}}=\hbar^2k^2/(2m)$ and
thus can be neglected if the characteristic interaction time between the two
dark-state polaritons $\tau_{\mathrm{pulse}}=l/v_{g1}$ is small compared with
the reciprocal recoil frequency:
$\omega_{\mathrm{rec}}\tau_{\mathrm{pulse}}\ll1$, where $l$ is the length of
the probe pulse in the medium. The latter condition can be easily fullfilled
for typical slow light pulses whose durations are of the order of a microsecond
\cite{Hau-99} and thus are much smaller than the reciprocal recoil frequencies.
Consequently the polaritons $\Phi_{D1}$ and $\Phi_{D2}$ are decoupled and
equations for them can be solved separately.

We are interested in the equation for the first dark polariton. Such a
polariton contains the radiative contribution and thus describes propagation of
the probe pulse of light in the medium. Neglecting the coupling with the second
polariton, Eq.~(\ref{eq:polar2}) yields a closed equation for the paraxial
propagation of the first polariton along the $z$ direction:
\begin{equation}
i\hbar\left(\frac{\partial}{\partial t}+v_{g1}\frac{\partial}{\partial
z}\right)\Phi_{D1}^{\prime}=-\frac{\hbar^2}{2m_{D1}}\nabla^2\Phi_{D1}^{\prime}
+i\hbar\mathbf{J}_{11}^{\prime}\cdot\nabla\Phi_{D1}^{\prime}+U_{11}^{
\prime}\Phi_{D1}^{\prime}\,,
\label{eq:decoupled-phiD1-para}
\end{equation}
with $v_{g1}\approx v_{\mathrm{rad}}$. Due to the finite lifetime of the
excited atomic state $\gamma^{-1}$, the first polariton will experience
radiative losses which are not included in the propagation equations
(\ref{eq:polar2}) and (\ref{eq:decoupled-phiD1-para}). Let us now estimate the
losses. The polariton lifetime is determined by the rate of the excited state
decay and the total Rabi frequency of the control lasers $\Omega_c$
\cite{Juzeliunas-02}:
$\tau_{\mathrm{pol}}=\gamma^{-1}(\Omega_c/\Delta\omega)^2$, where
$\Delta\omega$ is a detuning from the two-photon resonance. One of the reasons
for the appearance of the two-photon detunning is the finite duration of the
probe pulse, $\Delta\omega=\tau_{\mathrm{pulse}}^{-1}$. To avoid the losses, a
time the polariton tranverses the sample should be smaller than the polariton
lifetime: $L/v_{\mathrm{rad}}\ll\tau_{\mathrm{pol}}$, with $L$ being the length
of the atomic cloud. This means the total Rabi frequency $\Omega_c$ should be
large enough, 
\begin{equation}
L\ll
v_{\mathrm{rad}}\gamma^{-1}\Omega_c^2\tau_{\mathrm{pulse}}^2\,.
\label{eq:Omega-c-condition}
\end{equation}
Note also that in the slow light regime, the probe radiation makes a tiny
contribution to the polariton which is composed predominantly of the atomic
excitations (atomic coherences). In fact, the velocity ratio
$v_{\mathrm{rad}}/c\ll1$ represents a fraction of the radiation component in
the polariton \cite{Fleischhauer-00,Juzeliunas-02}. Thus
Eq.~(\ref{eq:decoupled-phiD1-para}) effectively describes propagation of the
atomic coherences along the $z$ axis at the velocity $v_{\mathrm{rad}}\ll c$
appearing due to the small radiative component.

\section{Storage and release the slow light:
General\label{sec:Storage-and-release}}

\subsection{Storage of slow light}

Let us first consider the storage of the slow light. The probe beam
$\mathcal{E}^{(s)}$ enters the atomic medium at $z=z_0$. The medium is
illuminated by two control beams characterized by Rabi frequencies
$\Omega_{c2}^{(\mathrm{s})}$ and $\Omega_{c3}^{(\mathrm{s})}$, where the index
$(\mathrm{s})$ refers to the storing stage of light. Initially the Rabi
frequencies of the control beams (and hence the group velocity $v_{g1}\equiv
v_{g1}^{(s)}$) are time-independent. Neglecting the diffraction effects, one
can thus write 
\begin{equation}
\mathcal{E}^{(s)}(t,z)=\mathcal{E}^{(s)}(\tau,z_0)\,,\qquad\tau=t-\int_{
z_0}^z\left(1/v_{g1}^{(s)}\right)dz^{\prime}.
\label{eq:E-s}
\end{equation}
At the boundary the probe beam is converted into a dark-state polariton
$\Phi_{D1}^{(\mathrm{s})}(t)$ propagating at the group velocity
$v_{g1}^{(s)}\ll c$ in the medium. Since the atomic population is created
exclusively by the incident probe light, only the first dark-state polariton is
populated, giving 
\begin{equation}
\Phi_{D1}^{(\mathrm{s})}=\mathcal{E}^{(s)}/\zeta_c^{(s)}\,,\qquad\Phi_{D2}^{
(\mathrm{s})}=0\,,
\label{eq:phi-D2-D1-initial}
\end{equation}
where the temporal and spatial dependence of the first polariton are kept
implicit. In writing the last relationship the use has been made of
Eq.~(\ref{eq:electric-DB}) relating $\mathcal{E}$ to $\Phi_{D1}$ and $\Phi_B$,
together with the adiabatic approximation implying that $\Phi_B\approx0$. For
slow light the parameter $\zeta_c^{(\mathrm{s})}\approx\sqrt{v_{g1}^{(s)}/c}$
featured in Eq.~(\ref{eq:phi-D2-D1-initial}) is much smaller than the unity.
That's why the dark-state polariton $\Phi_{D1}^{(\mathrm{s})}$ contains only a
tiny contribution by the electric field.

The equations~(\ref{eq:phi2-DB})--(\ref{eq:phi3-DB}) together with the
condition $\Phi_B^{(s)}(t)=\Phi_{D2}^{(\mathrm{s})}(t)=0$ provide the atomic
fields (atomic coherences) associated with the first polariton:
\begin{equation}
\Phi_2^{(\mathrm{s})}=-\xi_{c2}^{(\mathrm{s})*}\zeta_1^{(\mathrm{s})}\Phi_{D1}^{
(\mathrm{s})}\,,\qquad\Phi_3^{(\mathrm{s})}=-\xi_{c3}^{(\mathrm{s})*}\zeta_1^{
(\mathrm{s})}\Phi_{D1}^{(\mathrm{s})}\,.
\label{eq:phi-2--phi-3}
\end{equation}
At a certain time $t=t^{(s)}$ the whole probe pulse enters the atomic medium
and is contained in it. To store the slow polariton, both control fields are
switched off at $t=t^{(s)}$ in such a way that the Rabi frequency ratio
$\Omega_{c2}^{(\mathrm{s})}/\Omega_{c3}^{(\mathrm{s})}=\xi_{c2}^{(\mathrm{
s})}/\xi_{c3}^{(\mathrm{s})}$ remains constant, whereas
$\zeta_1^{(\mathrm{s})}\rightarrow1$. This gives the following atomic fields
(atomic coherences) at the storing time:
\begin{equation}
\Phi_2^{(\mathrm{s})}(t^{(s)})\rightarrow-\xi_{c2}^{(\mathrm{s})*}\Phi_{D1}^{
(\mathrm{s})}(t^{(s)})\,,\qquad\Phi_3^{(\mathrm{s})}(t^{(s)})\rightarrow-\xi_{
c3}^{(\mathrm{s})*}\Phi_{D1}^{(\mathrm{s})}(t^{(s)})\,.
\label{eq:phi23-final}
\end{equation}
The stored atomic coherences no longer have the radiative group velocity and
thus are trapped in the medium. The retrieval of these coherences is
accomplished at a later time $t=t^{(r)}$.

\subsection{Regeneration of slow light}

To restore the polariton propagation, both control fields are switched on again
at $t=t^{(r)}$ in such a way that their the ratio
$\Omega_{c2}^{(\mathrm{r})}/\Omega_{c3}^{(\mathrm{r})}=\xi_{c2}^{(\mathrm{
r})}/\xi_{c3}^{(r)}$ is constant. The difference between the storage and the
retrieval times should not be too large, so that the atomic coherences given by
Eq.~(\ref{eq:phi23-final}) are preserved up to the retrieval time. In the
initial experiment \cite{Liu-01} the light was stored up to a millisecond, yet
the storage duration was increased up to a second recently
\cite{Bloch09PRL,Hau09PRL}.

If the relative Rabi frequencies $\xi_{c2}^{(r)}$ and $\xi_{c3}^{(r)}$ differ
from the original ones $\xi_{c2}^{(\mathrm{s})}$ and $\xi_{c3}^{(\mathrm{s})}$,
both dark-state polaritons are regenerated. Using Eqs.~(\ref{eq:darkP1}) and
(\ref{eq:darkP2}), the dark state polaritons regenerated from the atomic
coherences (\ref{eq:phi23-final}) read at the beginning of the release of light
where $\zeta_1^{(r)}\approx1$:
\begin{eqnarray}
\Phi_{D1}^{(r)}(t^{(r)}) & = &
(\xi_{c2}^{(r)}\xi_{c2}^{(\mathrm{s})*}+\xi_{c3}^{(r)}\xi_{c3}^{(\mathrm{
s})*})\Phi_{D1}^{(\mathrm{s})}(t^{(s)})\,,
\label{eq:phiD1-regenerated}
\\
\Phi_{D2}^{(r)}(t^{(r)}) & = &
-(\xi_{c3}^{(r)*}\xi_{c2}^{(\mathrm{s})*}-\xi_{c2}^{(r)*}\xi_{c3}^{(\mathrm{
s})*})\Phi_{D1}^{(\mathrm{s})}(t^{(s)})\,.
\label{eq:phiD2-regenerated}
\end{eqnarray}
The electric probe field reappears due to the first dark-state polariton
containing a non-zero electric field contribution: 
\begin{equation}
\mathcal{E}^{(r)}(t)=\zeta_c^{(r)}(t)\Phi_{D1}^{(r)}(t)\,.
\label{eq:electric-released}
\end{equation}
Substitution of Eq.~(\ref{eq:phiD1-regenerated}) into
Eq.~(\ref{eq:electric-released}) and using Eq.~(\ref{eq:phi-D2-D1-initial}),
one can relate the regenerated electric field to the initial one as 
\begin{equation}
\mathcal{E}^{(r)}(t^{(r)})=\frac{\zeta_c^{(r)}}{\zeta_c^{(s)}}(\xi_{c2}^{
(r)}\xi_{c2}^{(\mathrm{s})*}+\xi_{c3}^{(r)}\xi_{c3}^{(\mathrm{s})*})\mathcal{
E}^{(s)}(t^{(s)}).
\label{eq:electric-release-expanded}
\end{equation}
If both the storing and the retrieval takes place in the slow light regime,
$\Omega_c^{(s)}\ll g\sqrt{n}$ and $\Omega_c^{(r)}\ll g\sqrt{n}$ , the above
equation simplifies to
\begin{equation}
\mathcal{E}^{(r)}(t^{(r)})=\frac{\Omega_{c2}^{(r)}\Omega_{c2}^{(\mathrm{s})*}
+\Omega_{c3}^{(r)}\Omega_{c3}^{(\mathrm{s})*}}{\left|\Omega_{c2}^{(s)}\right|^2
+\left|\Omega_{c3}^{(s)}\right|^2}\mathcal{E}^{(s)}(t^{(s)}).
\label{eq:electric-release-expanded-altern}
\end{equation}
Propagation of the regenerated polariton $\Phi_{D1}^{(r)}$ is governed by
Eq.~(\ref{eq:decoupled-phiD1-para}) in the paraxial case. The polariton
$\Phi_{D1}^{(r)}$ propagates at the velocity $v_{g1}\equiv v_{g1}^{(r)}$ and
might experience diffraction effects due to the second order transverse
derivatives featured in Eq.~(\ref{eq:decoupled-phiD1-para}). On the other hand,
the second polariton $\Phi_{D2}$ is not coupled to the light fields and hence
remains trapped (frozen) in the medium.

\section{Storage and retrieval of slow light: Specific situations}

\subsection{Restored control beams with the same spatial behaviour}

Let us first analyze the simplest situation where the Rabi frequencies of the
restored control beams are proportional to the corresponding original ones with
the same proportionality constant $b$: 
\begin{equation}
\Omega_{c2}^{(r)}=b\Omega_{c2}^{(\mathrm{s})}\,,\qquad\Omega_{c3}^{
(r)}=b\Omega_{c3}^{(\mathrm{s})}
\label{eq:Omega-r-s--Case1}
\end{equation}
and hence $\xi_{c2}^{(r)}=\xi_{c2}^{(\mathrm{s})}$ and
$\xi_{c3}^{(r)}=\xi_{c3}^{(\mathrm{s})}$. Under these conditions,
Eqs.~(\ref{eq:phiD1-regenerated}) and (\ref{eq:phiD2-regenerated}) together
with (\ref{eq:zeta-2--3}) provide the following amplitudes of the regenerated
dark-state polaritons:
\begin{equation}
\Phi_{D1}^{(r)}(t^{(r)})=\Phi_{D1}^{(\mathrm{s})}(t^{(s)})\,,\qquad\Phi_{D2}^{
(r)}=0\,.
\label{eq:regenerated-fields-1}
\end{equation}
Thus the second polariton is not populated ($\Phi_{D2}^{(r)}=0$), whereas the
first regenerated dark-state polariton coincides with the original one. The
corresponding regenerated electric field 
\begin{equation}
\mathcal{E}^{(r)}=b\mathcal{E}^{(s)}
\label{eq:electric-release-case-1}
\end{equation}
is proportional to the original one and thus does not acquire the phase
singularity of the control beam $\Omega_{c2}$ (if any). In such a situation the
vortex can not be transferred from the control to the regenerated probe beam.
In the following Subsections we will analyse the vortex transfer from the
control beam $\Omega_{c2}$ to the regenerated probe beam in the case where the
condition (\ref{eq:Omega-r-s--Case1}) no longer holds. Such a vortex transfer
is accompanied with some population of the second polariton.

It is noteworthy that the regenerated electric field $\mathcal{E}^{(r)}$ given
by Eq.~(\ref{eq:electric-release-case-1}) is increased (decreased) if the ratio
of the total Rabi frequencies $b=\Omega_c^{(r)}/\Omega_c^{(s)}$ is larger
(smaller) than the unity. On the other hand the group velocity is increased for
$b>1$ and decreased for $b<1$. This leads to the compression (for $b<1$) or
decompression for ($b>1$) of the regenerated probe pulse as compared to the
stored one, a feature known from the light storage and retrieval in the
$\Lambda$ system \cite{Liu-01}. Note also that the total number of the
regenerated photons is the same as that in the input beam. This is because the
second polariton is not populated $\Phi_{D2}=0$, so no atomic coherence remains
frozen in the medium.

\subsection{Transfer of optical vortex at the retrieval of the probe beam}

Suppose that only one control field is used during the storage phase of the
probe light, i.e. $\Omega_{c3}^{(\mathrm{s})}=0$ and hence
$|\xi_{c2}^{(\mathrm{s})}|=1$. This means the storage stage involves a
$\Lambda$ scheme depicted in Fig.~\ref{fig:fig1-Lambda}. In such a setup, the
control beam $\Omega_{c2}^{(\mathrm{s})}$ can not carry an OAM: Otherwise there
would be non-adiabatic losses of the probe beam at the vortex core of the
control beam. On the other hand, the retrieval of the probe beam is
accomplished using a tripod system in which generally both $\Omega_{c2}^{(r)}$
and $\Omega_{c3}^{(r)}$ are non-zero. Under these conditions,
Eqs.~(\ref{eq:phiD1-regenerated}) and (\ref{eq:phiD2-regenerated}) provide the
following results for the regenerated polaritons:
\begin{eqnarray}
\Phi_{D1}^{(r)}(t_i^{(r)}) & = &
\xi_{c2}^{(r)}\xi_{c2}^{(\mathrm{s})*}\Phi_{D1}^{(\mathrm{s})}(t_f^{(s)})\,,
\label{eq:phiD1-regenerated-2}
\\\Phi_{D2}^{(r)}(t_i^{(r)}) & = &
-\xi_{c3}^{(r)*}\xi_{c2}^{(\mathrm{s})*}\Phi_{D1}^{(\mathrm{s})}(t_f^{(s)})\,.
\label{eq:phiD2-regenerated-2}
\end{eqnarray}
The second polariton given by Eq.~(\ref{eq:phiD2-regenerated-2}) does not have
a radiative component and is trapped in the medium. The electric field of the
probe beam is regenerated exclusively due to the first polariton and reads
using Eq.~(\ref{eq:electric-release-expanded-altern})
\begin{equation}
\mathcal{E}^{(r)}=\frac{\Omega_{c2}^{(r)}}{\Omega_{c2}^{(s)}}\mathcal{E}^{
(s)}(t^{(s)}).
\label{eq:E-r-trip-case-final}
\end{equation}
The equations (\ref{eq:phiD1-regenerated-2}) or (\ref{eq:E-r-trip-case-final})
represent the initial condition for the subsequent propagation of the polariton
in the medium. Such a polariton will propagate along the $z$ axis with the
group velocity $v_{\mathrm{rad}}$. Its transverse profile will change due to
the diffraction effects represented by the second order spatial derivatives in
Eq.~(\ref{eq:decoupled-phiD1-para}).

If the control beam $\Omega_{c2}$ carries an optical vortex at the retrieval
stage, $\Omega_{c2}^{(r)}\sim e^{i\ell\varphi}$, the regenerated electric field
$\mathcal{E}^{(r)}\sim e^{i\ell\varphi}$ acquires the same phase as one can see
from Eq.~(\ref{eq:E-r-trip-case-final}). This means that the restored control
beam transfers its optical vortex to the regenerated electric field
$\mathcal{E}^{(r)}$. If the initial control and probe fields have the same
transverse dependence, the transverse profile of the regenerated electric field
will mimic that of the control field
$\mathcal{E}^{(r)}\sim\Omega_{c2}^{(r)}\sim e^{i\ell\varphi}$.

As an illustration, let us take the restoring control laser $\Omega_{c2}^{(r)}$
to be the first order Laguerre-Gaussian (LG) beam:
$\Omega_{c2}^{(r)}=A\tilde{\rho}e^{i\varphi}\exp(-\tilde{\rho}^2/\sigma_r^2)$,
where $\tilde{\rho}=\rho/\lambda$ is a dimensionless cylindrical radius,
$\lambda=2\pi/k$ being the optical wave-length. On the other hand, the control
beam is assumed to be the zero-order LG beam during the storage stage involving
a $\Lambda$ system: $\Omega_{c2}^{(s)}=a^{-1}A\exp(-\tilde{\rho}^2/\sigma_s^2)$
, where $a$ determines a relative amplitude of the control fields
$\Omega_{c2}^{(r)}$ and $\Omega_{c2}^{(s)}$ , $\sigma_r$ and $\sigma_s$ being
their dimensionless widths. This provides the following regenerated probe field
\begin{equation}
\mathcal{E}^{(r)}=a\tilde{\rho}e^{i\varphi}\exp\left[-\tilde{
\rho}^2\left(\sigma_r^{-2}-\sigma_s^{-2}\right)\right]\mathcal{E}^{(s)}(t^{
(s)})\,.
\label{eq:E-r-lambda-storing}
\end{equation}
It is noteworthy that the Rabi frequency of the additional laser
$\Omega_{c3}^{(r)}$ does not enter the above equations
(\ref{eq:E-r-trip-case-final})--(\ref{eq:E-r-lambda-storing}) for the
regenerated probe field. Yet the additional laser plays an important role to
ensure the lossles propagation of the restored probe field in a vicinity of the
vortex core where $\Omega_{c2}^{(r)}\rightarrow0$, as one can see from
Eq.~(\ref{eq:Omega-c-condition}).

\subsection{Transfer of the optical vortex during the storage of slow light}

Consider next the opposite situation where both control fields are on during
the storage phase, so the storage of the probe beam is carried out using a
tripod scheme. On the other hand, a $\Lambda$ scheme is employed during the
retrieval of the probe beam where only one control field is on, i.e.\
$\Omega_{c3}^{(r)}=0$ and hence $|\xi_{c2}^{(r)}|=1$. In that case
Eqs.~(\ref{eq:phiD1-regenerated}) and (\ref{eq:phiD2-regenerated}) yield the
following results for the regenerated polaritons:
\begin{eqnarray}
\Phi_{D1}^{(r)}(t_i^{(r)}) & = &
\xi_{c2}^{(r)}\xi_{c2}^{(\mathrm{s})*}\Phi_{D1}^{(\mathrm{s})}(t_f^{(s)})\,,
\label{eq:phiD1-regenerated-3}
\\\Phi_{D2}^{(r)}(t_i^{(r)}) & = &
\xi_{c2}^{(r)*}\xi_{c3}^{(\mathrm{s})*}\Phi_{D1}^{(\mathrm{s})}(t_f^{(s)})\,.
\label{eq:phiD2-regenerated-3}
\end{eqnarray}
Again the electric probe field is regenerated exclusively due to the first
polariton and is given by using Eq.~(\ref{eq:electric-release-expanded-altern})
\begin{equation}
\mathcal{E}^{(r)}=\frac{\Omega_{c2}^{(r)}\Omega_{c2}^{(\mathrm{s})*}}{
\left|\Omega_{c2}^{(s)}\right|^2+\left|\Omega_{c3}^{(s)}\right|^2}\mathcal{E}^{
(s)}(t^{(s)}).
\label{eq:E-r-Lambda-case-final}
\end{equation}
The equations (\ref{eq:phiD1-regenerated-3}) or
(\ref{eq:E-r-Lambda-case-final}) represent the initial conditions for the
subsequent propagation of the regenerated polariton governed, in the paraxial
case, by the equation of motion (\ref{eq:decoupled-phiD1-para}). Such a
polariton will propagate along the $z$ axis with the group velocity
$v_{\mathrm{rad}}$, and its transverse profile will change due to the
diffraction effects represented by the second order spatial derivatives in
Eq.~(\ref{eq:decoupled-phiD1-para}). On the other hand, the second polariton
$\Phi_{D2}$ will be frozen in the medium (neglecting the atomic motion) and its
spatial form is given by Eq.~(\ref{eq:phiD2-regenerated-3}).

If the second control beam carries an optical vortex at the storing stage,
$\Omega_{c2}^{(s)}\sim e^{i\ell\varphi}$, the regenerated electric field
$\mathcal{E}\sim e^{-i\ell\varphi}$ acquires an opposite vorticity, as one can
see from Eqs.~(\ref{eq:phiD1-regenerated-3}) and
(\ref{eq:E-r-Lambda-case-final}). The additional control beam
$\Omega_{c3}^{(s)}$ does not have a vortex and hence is non-zero at the center.
This ensures the lossless (adiabatic) propagation of the probe beam during the
storage phase. It is noteworthy that the transverse profile of the regenerated
probe field differs now from that of the storing beam
$\Omega_{c2}^{(\mathrm{s})}$ due to the denominator in
Eq.~(\ref{eq:E-r-Lambda-case-final}).

Suppose that the control lasers are the first and zero order LG beams at the
storage stage: 
\begin{equation}
\Omega_{c2}^{(s)}=A\tilde{\rho}e^{i\varphi}\exp(-\tilde{
\rho}^2/\sigma_s^2)\,,\quad\Omega_{c3}^{(s)}=bA\exp(-\tilde{
\rho}^2/\sigma_s^2)\,,
\end{equation}
where the parameter $b$ determines the relative amplitude of the additional
control laser. On the other hand, the control beam is assumed to be the
zero-order LG beam at the retrieval stage involing the $\Lambda$ scheme:
$\Omega_{c2}^{(r)}=aA\exp(-\tilde{\rho}^2/\sigma_r^2)$. Thus one arrives at the
following regenerated probe field containing the phase conjugated vortex

\begin{equation}
\mathcal{E}^{(r)}=\frac{a}{\tilde{\rho}^2+b^2}\tilde{\rho}e^{
-i\varphi}\exp\left[-\tilde{\rho}^2\left(\sigma_r^{-2}-\sigma_s^{
-2}\right)\right]\mathcal{E}^{(s)}(t^{(s)})\,.
\label{eq:E-r-lambda-release}
\end{equation}
It is noteworthy that for the $b<1$ the transverse profile of the regenerated
beam can differ considerably from the the Laguerre-Gaussian shape. Decreasing
$b$ the transverse shape of the regenerated beam becomes narrower. This leads
to a larger difraction in its subsequent propagation, as it will be explored in
the following Subsection.

\subsection{Dynamics of the restored probe beams}

Let us suppose that the atomic cloud is small enough, so that the diffraction
can be neglected during the propagation of the probe beam in the medium. Such a
condition can be fulfilled readily for a typical cloud of cold atoms, the
length of which normally does not exceed a third of the millimeter
\cite{Hau-99}. After leaving the atomic cloud, the probe beam propagates in the
free space according to Eq.~(\ref{eq:electric}) with $g=0$. Since the probe
field is quasimonochromatic, its amplitude $\mathcal{E}(\mathbf{r},t)$ changes
little during an optical cycle. In the stationary case one arrives at the
following propagation equation for the slowly varying amplitude of the probe
field:
\begin{equation}
i\frac{\partial}{\partial
z}\mathcal{E}=-\frac{1}{2k}\nabla_{\bot}^2\mathcal{E}\,.
\label{eq:electric-free}
\end{equation}

In the previous two subsections we have considered two possible scenarios to
regenerate the probe field. In the first ($\Lambda$-T) case the Lambda scheme
is used for storing the probe field whereas the tripod setup is employed for
the regeneration. In the second (T-$\Lambda$) case, the tripod scheme is used
for storing the probe field whereas the Lambda setup is used for the
regeneration. In what follows we shall explore the subsequent propagation of
the probe beam. The regenerated fields given by
Eqs.~(\ref{eq:E-r-lambda-storing}) and (\ref{eq:E-r-lambda-release}) represent
the initial conditions for such a propagation. By taking the initial probe beam
to be Gaussian
$\mathcal{E}^{(s)}=\mathcal{E}_0^{(s)}\exp(-\tilde{\rho}^2/\sigma_p^2)$, the
regenerated fields read for both cases 
\begin{equation}
\mathcal{E}_{\Lambda-T}^{(r)}=a\mathcal{E}_0^{(s)}\tilde{\rho}e^{i\varphi}e^{
-\tilde{\rho}^2/\sigma^2}\,,\qquad\mathcal{E}_{T-\Lambda}^{(r)}=\frac{a}{\tilde{
\rho}^2+b^2}\mathcal{E}_0^{(s)}\tilde{\rho}e^{-i\varphi}e^{-\tilde{
\rho}^2/\sigma^2}\,,
\label{eq:E-r-init-cond}
\end{equation}
where $\sigma^{-2}=\sigma_p^{-2}+\sigma_r^{-2}-\sigma_s^{-2}$ determines the
width of the regenerated probe field measured in optical wavelength
$\lambda=2\pi/k$.

\begin{figure}
\includegraphics[width=0.6\textwidth]{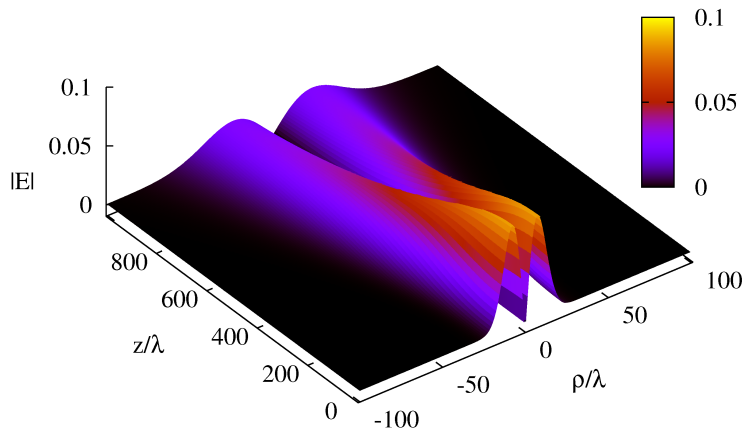}
\caption{Propagation of the regenerated probe field in the free space for
$\sigma=10$ and $a=1$. The $\Lambda$ scheme is used for storage and the tripod
system for retrieval of the probe field.}
\label{fig:fig2-1}
\end{figure}

The equation (\ref{eq:electric-free}) has been solved numerically for
$\sigma=10$. Figure~\ref{fig:fig2-1} shows the subsequent propagation of the
regenerated beam for the first case. Here the regenerated field
$\mathcal{E}_{\Lambda-T}^{(r)}$ represents the first order LG beam and is
proportional to the relative intensity of the control field at the release and
storage stages $\Omega_{c2}^{(r)}/\Omega_{c2}^{(s)}$. The subsequent
propagation of the field qualitatively preserves the transverse profile and is
accompanied by some the diffraction spreading.

\begin{figure}
\includegraphics[width=0.4\textwidth]{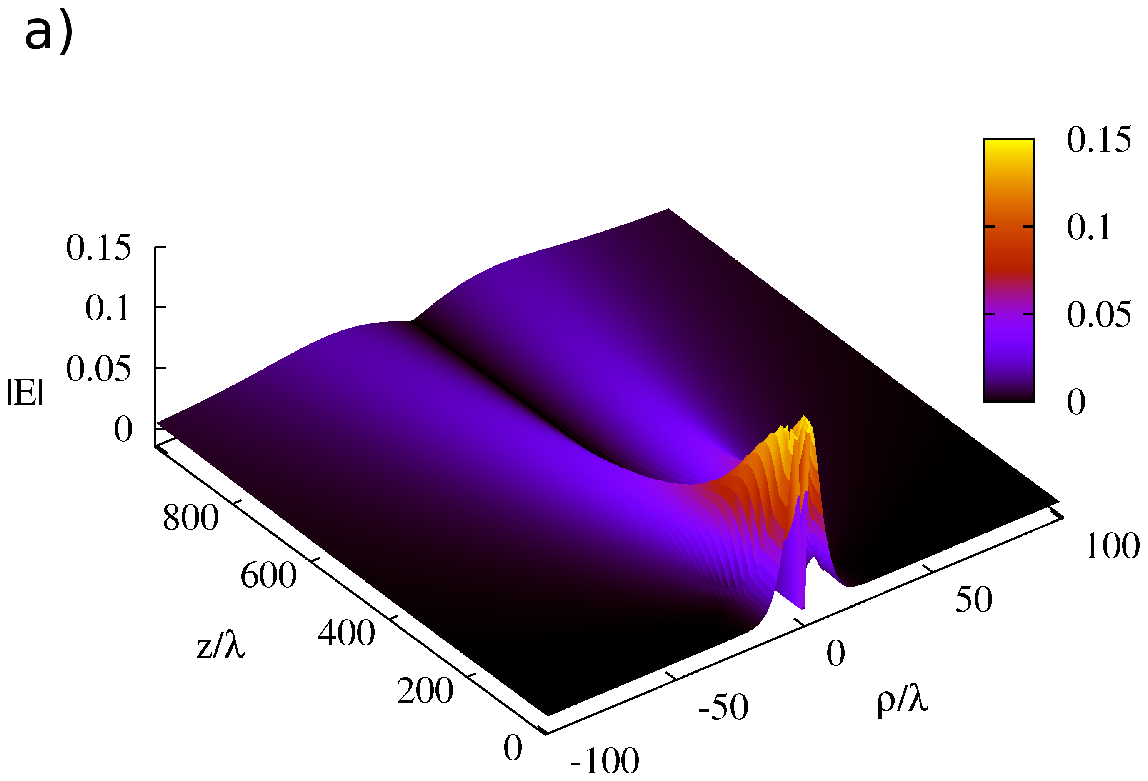}\includegraphics[width=0.4\textwidth]{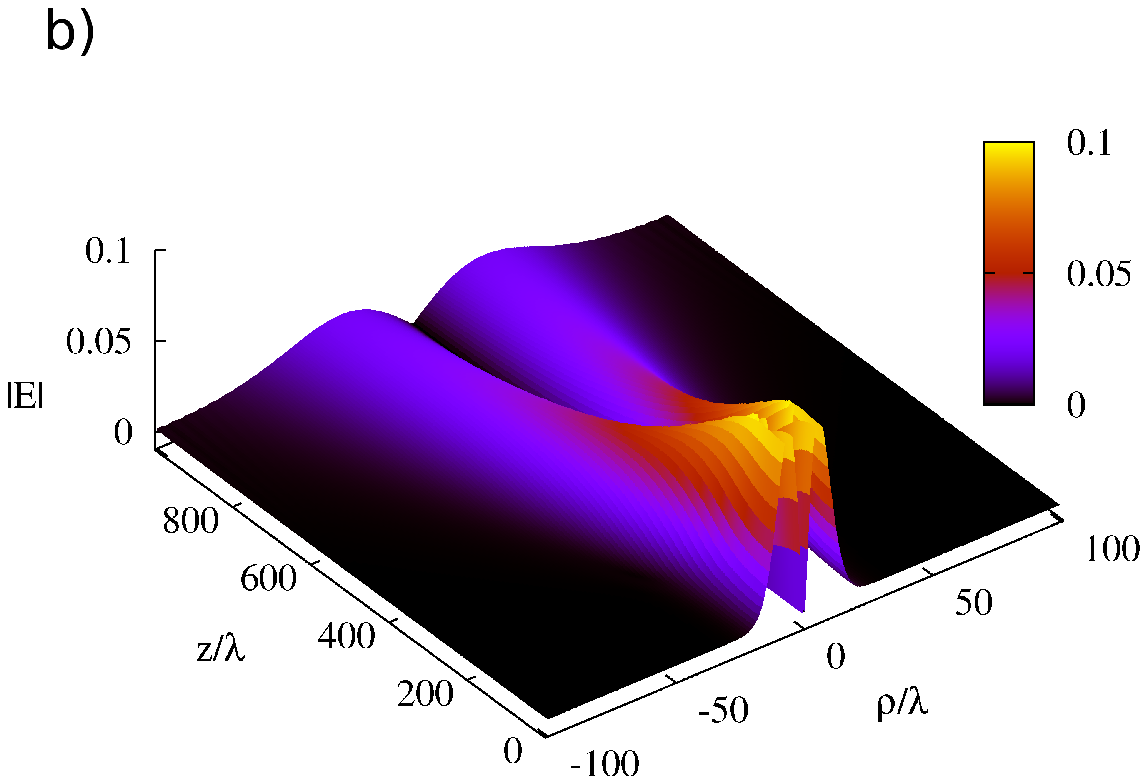}\newline\includegraphics[width=0.4\textwidth]{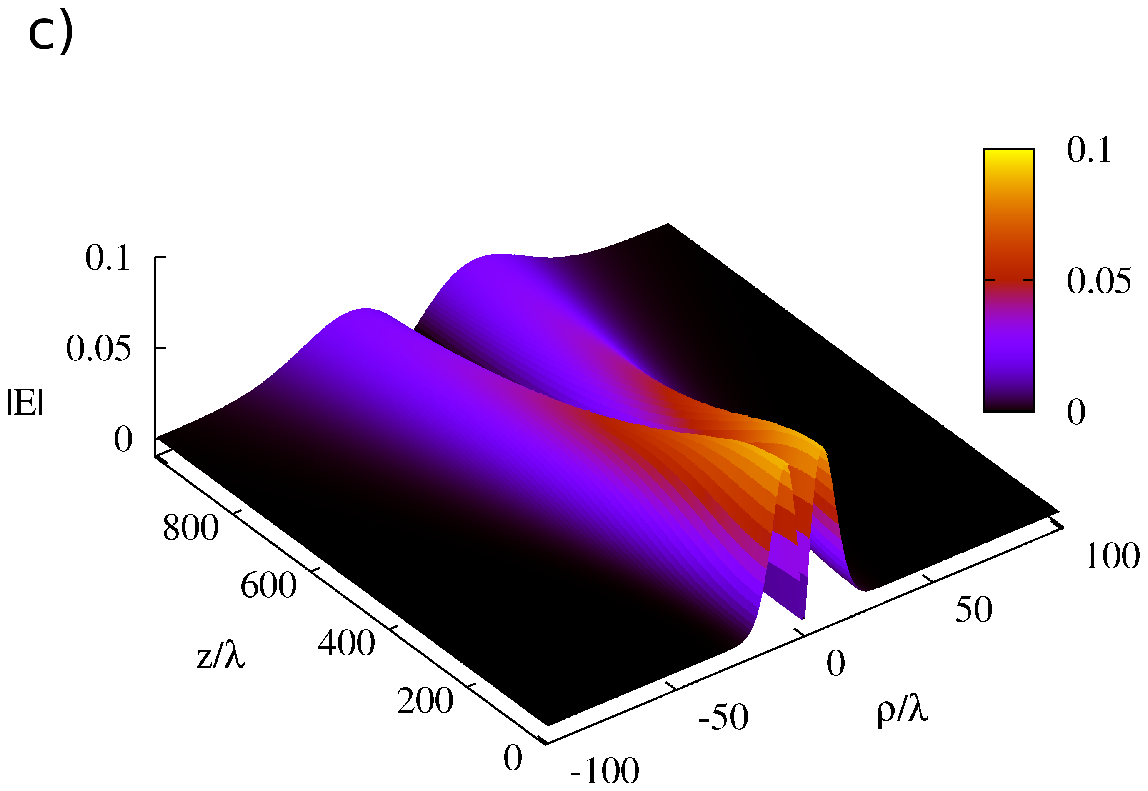}
\caption{Propagation of the regenerated electric field in the free space in
the case of tripod system for storage and $\Lambda$ system for retrieval by
taking $\sigma=10$ and $a=1$. The parameter $b$ appearing in
Eq.~(\ref{eq:E-r-init-cond}) is $b=3,\,10,\,30$ in the Figures a) b) and c),
respectively.}
\label{fig:fig3}
\end{figure}

The next three figures (Fig.~\ref{fig:fig3} a,b,c) illustrate the second case
where the tripod scheme is used for the storing and the $\Lambda$ scheme is
used for the retrieval of the probe beam. The transverse profile of the
regenerated beam $\mathcal{E}_{T-\Lambda}^{(r)}$ is determined by the relative
intensity $b$ of the additional laser beam $\Omega_{c3}^{(r)}$. For $b=3$ the
profile is much narrower as compared to the first order LG beam with the same
width $\sigma$, as one can see comparing Figs.~\ref{fig:fig2-1} and
\ref{fig:fig3}a. Consequently the light beam spreads out much faster than that
in the first case (see Fig.~\ref{fig:fig3}a). When $b$ increases ($b=10$ in
Fig.~\ref{fig:fig3}b and $b=30$ in Fig.~\ref{fig:fig3}c), the profile of the
probe beam approaches the shape featured in the first case. Note that the
increase in the relative intensity $b$ of the additional control laser is
accompanied by the decrease in the intensity of the regenerated probe beam.
Thus the improvement in the quality of the regenerated beams comes at a price
of reducing its intensity.

\section{Concluding remarks}

Polariton formalism has been applied for studying the propagation of a probe
field of light in a cloud of cold atoms influenced by two additional control
laser beams of larger intensity. The probe and control beams couple resonantly
three hyperfine ground states to a common excited state in a tripod
configuration of the atomic energy levels. The first control beam can have an
optical vortex. Application of another control beam without a vortex ensures
the loseless (adiabatic) propagation of the probe beam at the vortex core where
the intensity of the first control beam goes to zero. The adiabatic propagation
of the probe beam is obtained when the total intensity of the control lasers is
sufficiently large at the vortex core.

We have started with a set of atomic equations coupled with the equation for
the probe beam, subsequently transforming them into two coupled equations for
the dark-state polaritons. We have analysed conditions (related to the laser
pulse durations and switching times) when the polaritons are decoupled and thus
the problem reduces to a single equation for the polariton. An advantage of
polaritonic analysis is a simplicity of the relationship between the polariton
field and the regenerated electric field. Furthermore the equation for the
polariton has a usual form of matter wave equation which describes also the
atomic evolution when the control fields are off.

The probe pulse is stored onto the atomic coherences and subsequently retrieved
by switching off and on the control beams. As a result, the optical vortex can
be transferred from the control to the probe fields during the storage or
retrieval. Two scenarios have been analyzed in more details. The first case
involves a $\Lambda$ system for the storage and a tripod system for the
retrieval. In such a situation the phase of vortex is transferred from the
restoring control beam to the regenerated probe beam. In the second case the
tripod system is used for the storage and the $\Lambda$ system for the
retrieval. The vortex phase is then transferred from the storing control beam
to the regenerated probe beam in the phase conjugated form, so the probe beam
acquires an opposite vorticity. The profile of the regenerated probe field is
well preserved in the first case. On the other hand, in the second case the
regenerated beam becomes narrower and thus experience larger difraction
spreading. The width of the regenerated beam can be controlled by changing the
intensity of the additional control beam during the storage phase.

The tripod setup can be realized for atoms like Sodium \cite{Liu-01} 
or Rubidium  \cite{Phill2001} containing two hyperfine ground levels with $F=1$ and $F=2$,
as depicted in Fig.~\ref{fig:fig1}c. 
The scheme can be produced by adding an extra 
circularly polarized laser beam $\Omega_{c3}$ as compared to the experiment 
by Liu et al \cite{Liu-01} on the light storage in the  gases using the $\Lambda$ scheme. 
Thus it is feasible to implement the suggested experiment on the transfer 
of optical vortex from the control to the probe fields using the tripod setup.

\begin{acknowledgments}
The authors acknowledge the support by the Research Council of Lithuania
(Grants No. MOS-2/2010 and VP1-3.1-¶MM-01-V-01-001) and the EU FP7 project
STREP NAMEQUAM.
\end{acknowledgments}

\appendix

\section{Matrix elements in the equation for the dark polaritons}

\label{sec:appA}The elements of the matrix $\mathbf{J}$ featured in
Eq.~(\ref{eq:polar1}), are
\begin{eqnarray}
\mathbf{J}_{11} & = &
i\hbar\left(\frac{c^2}{\hbar\omega}\zeta_c\nabla\zeta_c+\frac{1}{
m}\zeta_1^*\nabla\zeta_1\right)+|\zeta_1|^2\mathbf{J}_{B2}\,,\\
\mathbf{J}_{22} & = &
i\frac{\hbar}{m}(\xi_{c3}^*\nabla\xi_{c3}+\xi_{c2}^*\nabla\xi_{c2})\,,\\
\mathbf{J}_{12} & = &
i\frac{\hbar}{m}\zeta_1^*(\xi_{c3}\nabla\xi_{c2}-\xi_{c2}\nabla\xi_{c3})\,,
\end{eqnarray}
where
\begin{equation}
\mathbf{J}_{B2}=i\frac{\hbar}{m}(\xi_{c2}\nabla\xi_{c2}^*+\xi_{c3}\nabla\xi_{
c3}^*)\,.
\end{equation}
The elements of the matrix $U$ in Eq.~(\ref{eq:polar1}), read
\begin{eqnarray}
U_{11} & = &
-\frac{\hbar^2}{2}\left(\frac{c^2}{\hbar\omega}\zeta_c\nabla^2\zeta_c+\frac{1}{
m}\zeta_1^*\nabla^2\zeta_1\right)+i\hbar\zeta_1^*\nabla\zeta_1\cdot\mathbf{J}_{
B2}\\ & &
+|\zeta_1|^2U_{B2}-\frac{\hbar\omega}{2}\zeta_c^2+i\hbar\left(\zeta_c\frac{
\partial}{\partial t}\zeta_c+\zeta_1\frac{\partial}{\partial
t}\zeta_1^*\right)\,,\nonumber\\ U_{22} & = &
-\frac{\hbar^2}{2m}(\xi_{c3}^*\nabla^2\xi_{c3}+\xi_{c2}^*\nabla^2\xi_{c2})\\ & &
+(\hbar\omega_{21}+V_2(\mathbf{r}))|\xi_{c3}|^2+(\hbar\omega_{31}+V_3(\mathbf{
r}))|\xi_{c2}|^2\nonumber\\ & &
+i\hbar\left(\xi_{c3}\frac{\partial}{\partial
t}\xi_{c3}^*+\xi_{c2}\frac{\partial}{\partial
t}\xi_{c2}^*\right)\,,\nonumber\\ U_{12} & = &
-\frac{\hbar^2}{2m}\zeta_1^*(\xi_{c3}\nabla^2\xi_{c2}-\xi_{c2}\nabla^2\xi_{
c3})\\ & &
+\zeta_1^*\xi_{c2}\xi_{c3}(\hbar\omega_{32}+V_3(\mathbf{r})-V_2(\mathbf{
r}))\nonumber\\ & & +i\hbar\zeta_1^*\left(\xi_{c2}\frac{\partial}{\partial
t}\xi_{c3}-\xi_{c3}\frac{\partial}{\partial t}\xi_{c2}\right)\,,\nonumber\\
U_{21} & = &
-\frac{\hbar^2}{2m}\zeta_1\left(\xi_{c2}^*\nabla^2\xi_{c3}^*-\xi_{
c3}^*\nabla^2\xi_{c2}^*\right)+i\hbar\frac{1}{\zeta_1}\nabla\zeta_1\cdot\mathbf{
J}_{21}\\ & &
+\zeta_1\xi_{c2}^*\xi_{c3}^*(\hbar\omega_{32}+V_3(\mathbf{r})-V_2(\mathbf{
r}))\nonumber\\ & &
+i\hbar\zeta_1\left(\xi_{c3}^*\frac{\partial}{\partial
t}\xi_{c2}^*-\xi_{c2}^*\frac{\partial}{\partial
t}\xi_{c3}^*\right)\,,\nonumber
\end{eqnarray}
where
\begin{eqnarray}
U_{B2} & = &
-\frac{\hbar^2}{2m}(\xi_{c2}\nabla^2\xi_{c2}^*+\xi_{c3}\nabla^2\xi_{c3}^*)\\ & &
+(\hbar\omega_{21}+V_2(\mathbf{r}))|\xi_{c2}|^2+(\hbar\omega_{31}+V_3(\mathbf{
r}))|\xi_{c3}|^2\nonumber\\ & &
+i\hbar\left(\xi_{c2}^*\frac{\partial}{\partial
t}\xi_{c2}+\xi_{c3}^*\frac{\partial}{\partial t}\xi_{c3}\right)\,.\nonumber
\end{eqnarray}

\section{Matrix elements in the paraxial equation for the dark polaritons}

\label{sec:appB}Using Eq.~(\ref{eq:Omega-c2--c3}), the parameters $\xi_{c2}$
and $\xi_{c3}$ have the form $\xi_{c2}=\xi_{c2}^{\prime}e^{ik_cz}$,
$\xi_{c3}=\xi_{c3}^{\prime}e^{ik_cz}$, where $\xi_{c2}^{\prime}$ and
$\xi_{c3}^{\prime}$ slowly change with the distance $z$ within the optical
wavelength. The diagonal elements of the matrix $\mathbf{J}^{\prime}$ entering
Eq.~(\ref{eq:polar2}), are given by
\begin{eqnarray}
\mathbf{J}_{11}^{\prime}& = &
i\hbar\left(\frac{c^2}{\hbar\omega}\zeta_c\nabla\zeta_c+\frac{1}{
m}\zeta_1^*\nabla\zeta_1\right)+|\zeta_1|^2\mathbf{J}_{B2}^{\prime}\,,\\
\mathbf{J}_{22}^{\prime}& = &
i\frac{\hbar}{m}(\xi_{c3}^{\prime*}\nabla\xi_{c3}^{\prime}+\xi_{c2}^{
\prime*}\nabla\xi_{c2}^{\prime})\,,
\end{eqnarray}
where
\begin{equation}
\mathbf{J}_{B2}^{\prime}=i\frac{\hbar}{m}(\xi_{c2}^{\prime}\nabla\xi_{c2}^{
\prime*}+\xi_{c3}^{\prime}\nabla\xi_{c3}^{\prime*})\,.
\end{equation}
The non-diagonal matrix elements of $\mathbf{J}^{\prime}$ read
\begin{eqnarray}
\mathbf{J}_{12}^{\prime}& = &
\mathbf{J}_{12}e^{-i(k+k_c)z}=i\frac{\hbar}{m}\zeta_1^*e^{i(k_c-k)z}(\xi_{c3}^{
\prime}\nabla\xi_{c2}^{\prime}-\xi_{c2}^{\prime}\nabla\xi_{c3}^{
\prime})\\\mathbf{J}_{21}^{\prime}& = &
\mathbf{J}_{21}e^{i(k+k_c)z}=\mathbf{J}_{12}^{\prime*}
\end{eqnarray}
The diagonal matrix elements of the complex scalar potential $U^{\prime}$ in
Eq.~(\ref{eq:polar2}), are
\begin{eqnarray}
U_{11}^{\prime}& = &
-\frac{\hbar^2}{2}\left(\frac{c^2}{\hbar\omega}\zeta_c\nabla^2\zeta_c+\frac{1}{
m}\zeta_1^*\nabla^2\zeta_1\right)+i\hbar\zeta_1^*\nabla\zeta_1\cdot\mathbf{J}_{
B2}^{\prime}\\ & &
+|\zeta_1|^2\left(U_{B2}^{\prime}+\frac{\hbar^2(k-k_c)^2}{2m}\right)
+i\hbar\left(\zeta_c\frac{\partial}{\partial
t}\zeta_c+\zeta_1\frac{\partial}{\partial
t}\zeta_1^*\right)\,,\nonumber\\ U_{22}^{\prime}& = &
-\frac{\hbar^2}{2m}(\xi_{c3}^{\prime*}\nabla^2\xi_{c3}^{\prime}+\xi_{c2}^{
\prime*}\nabla^2\xi_{c2}^{\prime})\\ & &
+(\hbar\omega_{21}+V_2(\mathbf{r}))|\xi_{c3}|^2+(\hbar\omega_{31}+V_3(\mathbf{
r}))|\xi_{c2}|^2\nonumber\\ & &
+i\hbar\left(\xi_{c3}^{\prime}\frac{\partial}{\partial
t}\xi_{c3}^{\prime*}+\xi_{c2}^{\prime}\frac{\partial}{\partial
t}\xi_{c2}^{\prime*}\right)\,,\nonumber
\end{eqnarray}
where
\begin{eqnarray}
U_{B2}^{\prime}& = &
-\frac{\hbar^2}{2m}(\xi_{c2}^{\prime}\nabla^2\xi_{c2}^{\prime*}+\xi_{c3}^{
\prime}\nabla^2\xi_{c3}^{\prime*})\\ & &
+(\hbar\omega_{21}+V_2(\mathbf{r}))|\xi_{c2}|^2+(\hbar\omega_{31}+V_3(\mathbf{
r}))|\xi_{c3}|^2\nonumber\\ & &
+i\hbar\left(\xi_{c2}^{\prime*}\frac{\partial}{\partial
t}\xi_{c2}^{\prime}+\xi_{c3}^{\prime*}\frac{\partial}{\partial
t}\xi_{c3}^{\prime}\right)\,.\nonumber
\end{eqnarray}
Finally, the non-diagonal elements of the complex scalar potential $U^{\prime}$
are given by
\begin{eqnarray*}
U_{12}^{\prime}& = &
-\frac{\hbar^2}{2m}\zeta_1^*e^{i(k_c-k)z}(\xi_{c3}^{\prime}\nabla^2\xi_{c2}^{
\prime}-\xi_{c2}^{\prime}\nabla^2\xi_{c3}^{\prime})\\ & &
+\zeta_1^*e^{i(k_c-k)z}\xi_{c2}^{\prime}\xi_{c3}^{\prime}(\hbar\omega_{32}
+V_3(\mathbf{r})-V_2(\mathbf{r}))\\ & &
+i\hbar\zeta_1^*e^{i(k_c-k)z}\left(\xi_{c2}^{\prime}\frac{\partial}{\partial
 t}\xi_{c3}^{\prime}-\xi_{c3}^{\prime}\frac{\partial}{\partial
t}\xi_{c2}^{\prime}\right)\,,\\ U_{21}^{\prime}& = &
-\frac{\hbar^2}{2m}\zeta_1e^{i(k-k_c)z}\left(\xi_{c2}^{\prime*}\nabla^2\xi_{
c3}^{\prime*}-\xi_{c3}^{\prime*}\nabla^2\xi_{c2}^{\prime*}\right)+i\hbar\frac{
1}{\zeta_1}\nabla\zeta_1\cdot\mathbf{J}_{21}^{\prime}\\ & &
+\zeta_1e^{i(k-k_c)z}\xi_{c2}^{\prime*}\xi_{c3}^{\prime*}(\hbar\omega_{32}
+V_3(\mathbf{r})-V_2(\mathbf{r}))\\ & &
+i\hbar\zeta_1e^{i(k-k_c)z}\left(\xi_{c3}^{\prime*}\frac{\partial}{\partial
t}\xi_{c2}^{\prime*}-\xi_{c2}^{\prime*}\frac{\partial}{\partial
t}\xi_{c3}^{\prime*}\right)\,.
\end{eqnarray*}

\end{document}